\documentclass[conference]{IEEEtran}
\IEEEoverridecommandlockouts
 
\usepackage{orcidlink} 
\usepackage{algorithm}
\usepackage{amsfonts}
\usepackage{hyperref}
\usepackage{cite}
\usepackage{algpseudocode}
\usepackage{graphicx}
\usepackage{textcomp}
\usepackage{lipsum}
\usepackage{tcolorbox}
\usepackage{soul}
\usepackage{xcolor}
\usepackage{tikz}
\usepackage{amsmath}
\usepackage{multirow}
\usepackage{cite}

\def\BibTeX{{\rm B\kern-.05em{\sc i\kern-.025em b}\kern-.08em
    T\kern-.1667em\lower.7ex\hbox{E}\kern-.125emX}}

\newcommand\encircle[1]{%
\tikz[baseline=(X.base)] 
 \node (X) [draw, scale=0.75, shape=circle, inner sep=0, fill=black, text=white, minimum size=0em] {\strut #1};}

\begin{document}

\title{
\vspace{-1.3cm}
{\fontsize{11}{16}\selectfont   \normalfont \textit{To Appear in the IEEE International Conference on Learning and Adaptive Systems (ICLAD), June 2025.}}
\texttt{FedChip}: Federated LLM for Artificial Intelligence Accelerator Chip Design}


\author{%
  Mahmoud Nazzal\orcidlink{0000-0003-3375-0310}$^{\S*}$,
  Khoa Nguyen\orcidlink{0009-0005-7605-7913}$^{\S*}$,
  Deepak Vungarala\orcidlink{0009-0000-8659-2040}$^{\S*}$,
  Ramtin Zand\orcidlink{0000-0002-1786-1152}$^{\P}$\\
  Shaahin~Angizi\orcidlink{0000-0003-2289-6381}$^{\S}$,
  Hai Phan\orcidlink{0000-0002-1032-8275}$^{\S}$,
  Abdallah Khreishah\orcidlink{0000-0003-1583-713X}$^{\S}$\\
\small
  $^\S$New Jersey Institute of Technology, Newark, NJ, USA, 
  $^\P$University of South Carolina, Columbia, SC, USA\\
  E-mail: \{mn69, nk569, dv336\}@njit.edu, ramtin@cse.sc.edu, \{shaahin.angizi, hai.phan, abdallah.khreishah\}@njit.edu\\
  $^*$These authors contributed equally.\\
  \vspace{-3em}
}

\maketitle

\begin{abstract}
AI hardware design is advancing rapidly, driven by the promise of design automation to make chip development faster, more efficient, and more accessible to a wide range of users. Amongst automation tools, Large Language Models (LLMs) offer a promising solution by automating and streamlining parts of the design process. However, their potential is hindered by data privacy concerns and the lack of domain-specific training. To address this, we introduce \texttt{FedChip}, a \underline{Fed}erated fine-tuning approach that enables multiple \underline{Chip} design parties to collaboratively enhance a shared LLM dedicated for automated hardware design generation while protecting proprietary data. \texttt{FedChip} enables parties to train the model on proprietary local data and improve the shared LLM's performance. To exemplify \texttt{FedChip}'s deployment, we create and release \texttt{APTPU-Gen}, a dataset of 30k design variations spanning various performance metric values such as power, performance, and area (PPA). To encourage the LLM to generate designs that achieve a balance across multiple quality metrics, we propose a new design evaluation metric, \textit{Chip@k}, which statistically evaluates the quality of generated designs against predefined acceptance criteria. Experimental results show that \texttt{FedChip} improves design quality by more than 77\% over high-end LLMs while maintaining data privacy.\footnote{Implementation and dataset are at \url{https://github.com/ACADLab/FedChip}.}
\end{abstract}



\section{Introduction}

\par The global Artificial Intelligence (AI) hardware market, fueled by innovations in deep learning accelerators, autonomous systems, and data-intensive applications, is expected to be worth \$84.9 billion by 2031 \cite{skyquest2023aihardware}. Despite this growth, escalating design complexity and the need for specialized expertise \cite{ren2023survey,genc2021gemmini} continue to pose significant challenges, making the process costly, time-intensive, and difficult to optimize for performance. To address these challenges, AI-based advancements in automated hardware design generation \cite{thakur2023verigen,he2023chateda,chang2023chipgpt,thakur2023autochip} are emerging as promising solutions, enabling faster, more cost-effective, and performance-optimized development.

\par Among AI models, Large Language Models (LLMs) have a promising potential in automated hardware design generation due to their attractive abilities to incorporate domain-specific knowledge and enable human-level interaction with design workflows \cite{chang2023chipgpt,10323953,he2023chateda,vungarala2024sa}. These abilities promise to open the design process to a broader range of developers. However, state-of-the-art LLMs such as OpenAI’s GPT-4o \cite{openai2024gpt4o} and Anthropic's Claude 3.5 Sonnet \cite{anthropic2024claude3.5} struggle to consistently generate practical hardware designs. Common flaws in their outputs include hallucinated variables, particularly for small-scale circuits, and an inability to handle complex AI accelerators effectively \cite{bommasani2021opportunities}.

\par The limitations of LLMs in hardware design stem from two key challenges: \ul{First}, the lack of access to sufficient domain-specific training data due to privacy and Intellectual Property (IP) restrictions imposed by companies like NVIDIA, AMD, and Intel \cite{xu2024llm,chen2024dawn}. Concerns over data breaches and competitive risks further discourage data sharing \cite{arstechnica2022nvidiacyberattack}. \ul{Second}, the complex trade-offs required in hardware design—balancing the power, performance, and area (PPA) metrics \cite{davis2021fast}—are highly workload-specific and demand detailed architectural expertise. Current LLMs lack domain-specific knowledge and precise control mechanisms to produce synthesizable and optimized designs. Addressing these challenges requires a collaborative framework that ensures data privacy while enabling specialized hardware design capabilities.

\begin{figure}[t]
\centering
\includegraphics[width=0.93\linewidth]{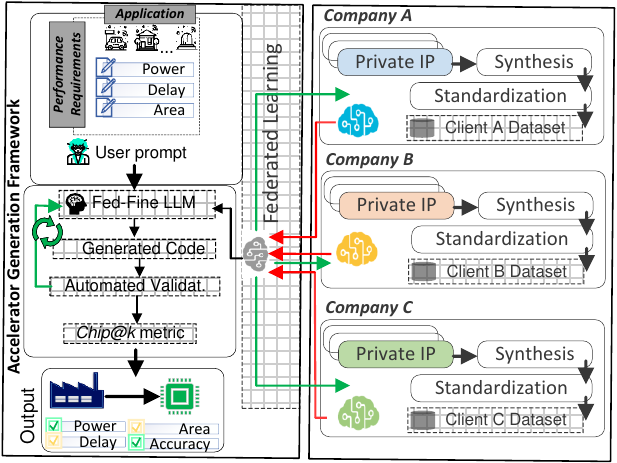}
\vspace{-1.1em}
\caption{\texttt{FedChip} enables multi-party LLM fine-tuning for chip design.}
\label{fig0}\vspace{-1em}
\end{figure}

\par Federated learning (FL) \cite{yang2019federated} trains a shared model across multiple clients without sharing raw data, thereby promoting data privacy \cite{bonawitz2019towards, mcmahan2017communication}. Therefore, FL is a promising framework for multi-party LLM training targeting automated hardware design generation. Recently, FL has been explored for fine-tuning LLMs across diverse tasks \cite{bai2024federated,kuang2024federatedscope,kuo2024federated}, focusing on reducing computational overhead and handling clients with different tasks. However, existing methods are not directly applicable to automated hardware design generation, where generating descriptive text for designs involves balancing multiple, often conflicting \cite{shi2020learned} objectives such as PPA metrics. There is an urgent need for a novel FL-based LLM fine-tuning approach tailored to the unique requirements of automated hardware design generation.

\par \textbf{Contributions.} Motivated by the above discussion, we propose \texttt{FedChip} illustrated in Fig. \ref{fig0}, a novel FL-based fine-tuning framework designed for collaborative automated hardware design generation across multiple manufacturers. \texttt{FedChip} addresses the conflicting goals \cite{shi2020learned} inherent in hardware design by enabling privacy-preserving collaboration while optimizing PPA metrics. 
Our key contributions are:
\begin{list}{$\bullet$}{\leftmargin=0em \itemindent=1em}
\item \textbf{Curated Dataset}: We introduce \texttt{APTPU-Gen}, a comprehensive dataset containing 30k hardware design variations annotated with PPA metrics, specifically designed to enhance LLM-based automated hardware design generation. The dataset is further clustered into sub-datasets reflecting the unique focuses of multiple hardware manufacturers, enabling realistic multi-party FL settings.

\item \textbf{A New FL Setting}: We design and implement \texttt{FedChip}, a privacy-preserving FL framework that allows multiple hardware manufacturers to collaboratively fine-tune LLMs for domain-specific tasks without compromising proprietary data or intellectual property.

\item \textbf{Evaluation Metric}: We propose a novel metric, \textit{Chip@k}, to statistically evaluate generated hardware designs based on their adherence to predefined PPA quality criteria, promoting balanced optimization across multiple design parameters.

\item \textbf{Empirical Validation}: Experimental results validating the effectiveness of the proposed \texttt{FedChip} in addressing privacy and hardware optimization challenges. \texttt{FedChip} is shown to improve design quality by more than 77\% over high-end LLMs while preserving data privacy and addressing key trade-offs in hardware design.
\end{list}

\section{Background}
\label{section2}
\par \textbf{LLMs for Hardware Design-Hardware Description is Text.} LLMs, as state-of-the-art text-generative models, are naturally suited to automated hardware design generation tasks since hardware design representations like Hardware Description Language (HDL) and High-Level Synthesis (HLS) code are textual. This property has led to efforts to employ LLMs to automate and improve hardware workflows. Along this line, LLMs have been applied to automate hardware design at different abstraction levels, such as RTL \cite{thakur2023verigen, he2023chateda} and Verilog HDL \cite{chang2023chipgpt,thakur2023autochip}. Furthermore, frameworks like GPT4AIGChip \cite{10323953} and SA-DS \cite{vungarala2024sa} explore domain-specific applications, including accelerator design and dataset creation. Despite these advancements, current frameworks face fundamental challenges such as limited domain-specific datasets and the absence of fine-tuning, which can result in hallucinated outputs and suboptimal performance on complex tasks \cite{he2023chateda,vungarala2024sa}. Fine-tuning, which can significantly improve LLM performance, is not yet widely applied in automated hardware design generation \cite{dai2022can}.

\par \textbf{LLMs and Fine-tuning.} Fine-tuning adapts general-purpose LLMs to domain-specific tasks by training on labeled datasets, enabling them to capture specialized patterns while maintaining their broad linguistic capabilities. This process improves task-specific performance metrics like accuracy and F1 score \cite{brown2020language,devlin2019bertpretrainingdeepbidirectional,liu2019roberta}. Furthermore, methods like low-rank adaptation (LoRA) and quantized LoRA (QLoRA) improve fine-tuning efficiency by selectively training a subset of model parameters while approximating the remaining ones, avoiding the need for full model retraining \cite{hu2021lora,dettmers2024qlora}. Hence, LLM fine-tuning is feasible on resource-constrained systems. However, the effectiveness of fine-tuning depends on high-quality domain data and robust evaluation metrics to ensure real-world applicability \cite{zhang2022opt}. This emphasizes the need to develop context-aware quality metrics for hardware design using LLMs.

\par \textbf{Federated Learning and Fine-Tuning of LLMs.} FL \cite{yang2019federated} is a decentralized machine learning approach where multiple clients collaboratively update a common model without sharing raw data, promoting data privacy. Each client trains the model locally and sends only model updates to a central server, aggregating these updates to improve the common model before redistributing it. FL reduces communication overhead and simplifies server-side training to an aggregation task, typically using methods like Federated Averaging (FedAvg), where client updates are weighted by dataset size \cite{mcmahan2017communication}. 

\begin{figure}[t]
\centering
\includegraphics[width=0.87\linewidth]{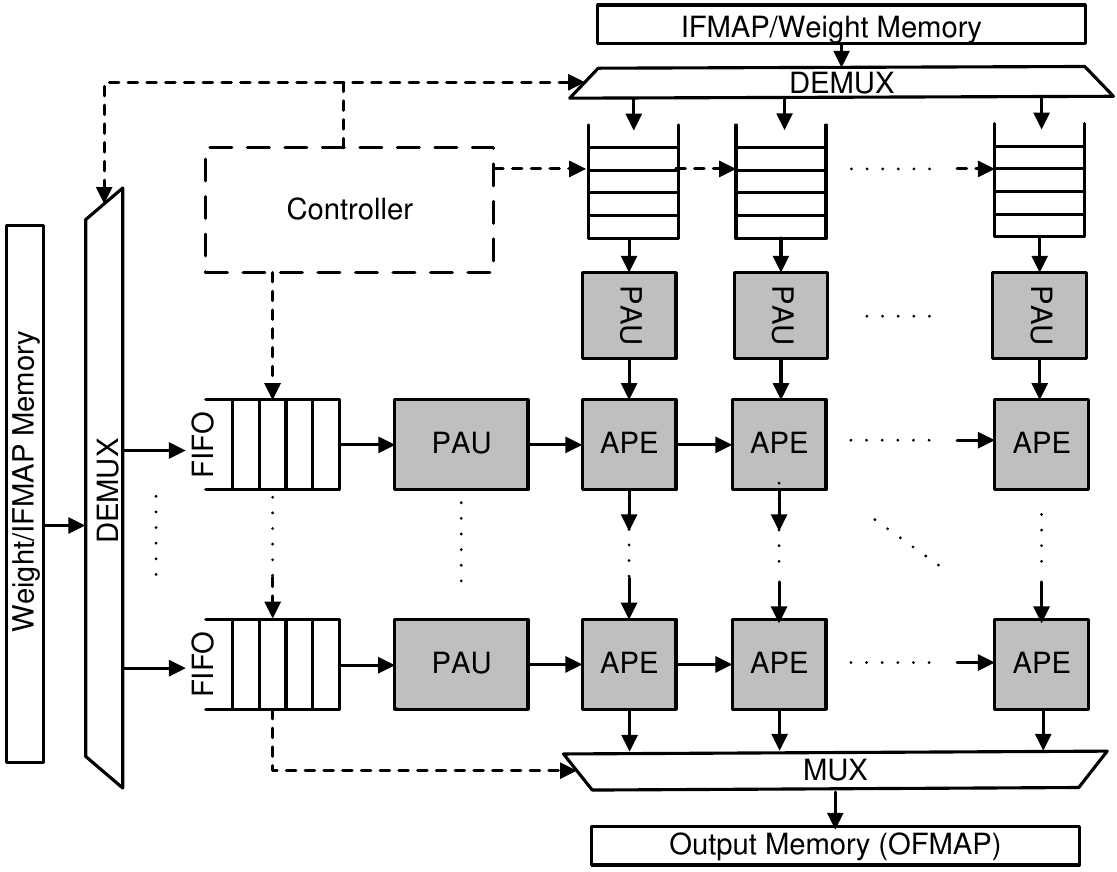}
\vspace{-1em}
\caption{The overall architecture of APTPU design.}
\label{tpuArch}
\vspace{-1.3em}
\end{figure}

\par Federated fine-tuning extends FL by adapting LLMs to domain-specific tasks using data distributed across multiple clients. Parameter-efficient techniques like LoRA have reduced computation and communication costs, allowing resource-constrained clients to participate effectively \cite{hu2021lora}. However, challenges persist, including task heterogeneity, resource disparities, and maintaining generalizability when aggregating updates from diverse domains. Recent methods like FlexLoRA address these issues by customizing LoRA ranks based on client resources, improving aggregation efficiency, and enabling more generalized global models \cite{bai2024federated, kuang2024federatedscope}. Despite these advancements, federated fine-tuning of LLMs still faces scalability issues, computational overhead, and the complexities of closed-source models. Additionally, current approaches are not tailored to the unique demands of hardware design, such as balancing multiple conflicting objectives (e.g., PPA metrics) and generating synthesizable, high-quality designs. These gaps highlight the need for specialized frameworks in this domain.

\par \textbf{APTPU.} The Approximate Tensor Processing Unit (APTPU) is a systolic array architecture designed to enhance performance and energy efficiency for deep learning workloads. It executes matrix-matrix, vector-vector, and matrix-vector multiplications by reusing data from memory, reducing buffer operations, and optimizing data flow \cite{TPU-micro2018}. A general sketch of APTPU is shown in Fig. \ref{tpuArch}.

\par As illustrated in Fig. \ref{tpuArch}, APTPU utilizes an Output Stationary (OS) data flow and features key components: weight/Input Feature Map (IFMap) memory, First-In, First-Out buffers (FIFOs), a controller, Pre-Approximate Units (PAUs), and Approximate Processing Elements (APEs). The controller manages data transfers, while PAUs dynamically adjust precision to balance accuracy and resource use. This design minimizes critical path delay and resource overhead, enabling low-precision Multiply-Accumulate (MAC) operations that enhance efficiency \cite{9901385}. APTPU's modular architecture highlights the trade-offs between PPA values, making it a valuable benchmark for evaluating LLM-based and FL-driven hardware optimization frameworks.

\par These trade-offs are captured through the PPA metrics involving power, performance (timing slack), and area metrics. Power refers to the total energy the design consumes during operation, where lower values indicate energy efficiency. Performance, typically evaluated through timing slack, measures whether the design meets its timing constraints; positive slack indicates success, while negative slack implies a timing violation. The area represents the physical silicon space occupied by design, with smaller areas being more cost-effective and efficient. These PPA metrics provide a holistic evaluation of hardware design quality and are integral to assessing APTPU's suitability for optimization tasks.


\section{Dataset Curation}
\label{section3}

\par In this work, we generate a dataset \texttt{APTPU-Gen} based on APTPU. Specifically, each data point in this dataset is composed of a \textit{verbal description} or an \textit{instruction} of a design and the design itself in Verilog HDL. APTPU is an ideal basis for dataset generation because it supports diverse hardware design configurations and deep learning workloads through its systolic array design. This makes it suitable for benchmarking designs across key performance metrics. APTPU designs are easy to analyze using third-party tools like the OpenROAD suite \cite{OPENROAD}, which provides a precise and standardized evaluation of physical parameters. Its modular and parameterized nature also allows the efficient generation of many design variations, making it ideal for exploring multi-objective optimization in FL settings.

\par
\begin{figure}[t] 
 \centering\includegraphics[width=0.9\linewidth]{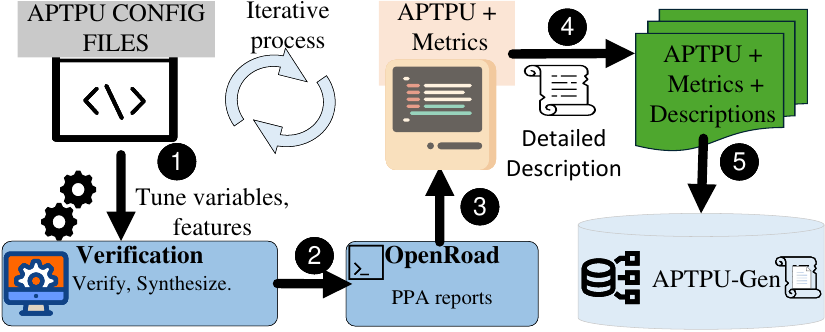} \vspace{-0.8em}
 \caption{The \texttt{APTPU-Gen} dataset curation.} \vspace{-1.2em}
 \label{aptpu}
\end{figure}

 The data curation process is shown in Fig.~\ref{aptpu}. We use the parameterized RTL of APTPU—hand-written in SystemVerilog with compile-time parameters for array dimensions, data widths, and approximation modes—to generate diverse architectural configurations (Step \encircle{1}). Unlike external code generators or LINT tools, this allows direct structural customization within the RTL itself. Next, we feed the generated designs to synthesis and functional verification (Step \encircle{2}). After that, the three PPA metrics of the generated design are evaluated using the OpenROAD suite \cite{OPENROAD} (Step \encircle{3}). Subsequently, the design with metrics is parsed using a script (Step \encircle{4}), resulting in the creation of a detailed, multi-level dataset that captures the reported PPA metrics (Step \encircle{5}). It is noted that Steps \encircle{1} to \encircle{3} are repeated iteratively multiple times until the desired architectural variations are generated. The time required for each data point generation varies depending on the specific configuration. To efficiently populate the \texttt{APTPU-Gen} dataset, we automate the generation of data points across different systolic array sizes, ensuring comprehensive coverage of design space exploration.

\par The \texttt{APTPU-Gen} dataset captures a comprehensive range of architectural configurations, comprising 29,952 variations for each of eight fixed systolic array implementations. These implementations correspond to compile-time array dimensions (e.g., 4$\times$4, 8$\times$8, \ldots, 256$\times$256) specified in the SystemVerilog RTL. The variations are generated by sweeping other compile-time parameters such as data width, approximation mode, and memory tiling. These configurations support workloads ranging from lightweight operations to large-scale deep neural network (DNN) computations. The core does not support arbitrary resizing or runtime scheduling across smaller multiplier units.

\begin{figure*}[t!]
 \centering
 \resizebox{0.96\linewidth}{!}{\includegraphics{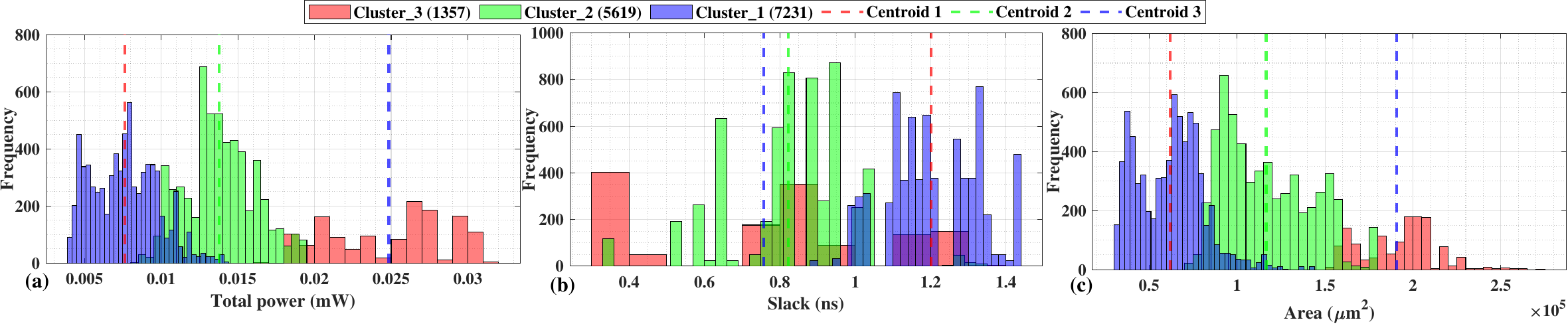}}
 \vspace{-1.2em}
\caption{Empirical distributions: (a) Total Power distribution across clusters, (b) Slack across clusters, and (c) Area distribution across clusters. Each distribution reflects the design priorities of one of the three emulated companies.}
 \label{partyDistributions} \vspace{-1em}
\end{figure*}

\par As of this work, \texttt{APTPU-Gen} includes 30k verified designs. Each design includes a top-level module summarizing the full circuit, suitable for tasks like retrieval-augmented generation (RAG). Although all entries stem from the APTPU core template, the dataset spans a broad design space by varying array dimensions, data widths, approximation modes, and memory-tiling strategies. This yields structurally and behaviorally diverse designs within the APTPU family. The dataset supports automated hardware generation, intelligent integration, and reuse across a variety of architectural settings. Extending the benchmark to other accelerator templates is left to future work.

\par Although some redundancy is inherent in the dataset due to the reuse of similar modules for different configurations, this structure allows for efficient exploration and customization of APTPU designs. The inclusion of detailed PPA metrics for each design iteration enhances the ability to generate budget-constrained designs and supports the development of optimized design space exploration strategies, accelerating the automated hardware design generation process.

\section{The Proposed \texttt{FedChip} Approach}
\label{section4}

\par Fig.~\ref{fig0} shows the \texttt{FedChip} pipeline, which combines FL and fine-tuned LLMs for collaborative chip design. On the left, a user prompt initiates the accelerator generation workflow; the LLM generates hardware designs, validated via the \textit{Chip@k} metric. On the right, multiple parties (A, B, C) fine-tune the LLM locally using their own data, sharing only model updates. These updates are then aggregated to improve the global model while preserving privacy.

\subsection{Modeling Chip Design Generation Across Multiple Parties}

\par \textbf{Data Splitting.} We exemplify the \texttt{FedChip} system model with three competitive but collaborating parties. Each party is a hardware accelerator design manufacturer. To emulate the operations of these parties, we cluster the 30k-sample \texttt{APTPU-Gen} dataset into three sub-datasets. Each sub-dataset is intended to represent the products from one of these parties. The clustering process is executed in two main steps. Initially, the dataset is normalized, and K-means clustering \cite{lloyd1982least} is applied to partition the data points into three clusters based on their PPA metrics. The K-means algorithm, known for its efficiency in minimizing within-cluster variance, is used to identify the core clusters characterizing each manufacturer's design focus. To introduce realistic variability and better reflect differences in business strategies or optimization focuses among the companies, we further refine the clustering using a stochastic Dirichlet sampling \cite{minka2000estimating}. Specifically, 20\% of the data points are randomly selected and reassigned to clusters based on probabilities drawn from a Dirichlet distribution \cite{minka2000estimating}, while the remaining 80\% of the data points retain their original cluster assignments. This ensures that while each manufacturer’s sub-dataset maintains the overall structure established by K-means, it also possesses a unique composition that reflects potential market or design variations.

\par Fig. \ref{partyDistributions}(a) presents the distribution of Total Power. The figure shows that Cluster 1 is characterized by the lowest power consumption, suggesting a focus on energy efficiency. Cluster 3, in contrast, displays the highest power consumption, indicating that designs in this cluster may prioritize performance or complexity over power efficiency. Fig. \ref{partyDistributions}(b) illustrates the distribution of Slack, where Cluster 1 tends to have the least negative slack, indicating tighter timing constraints and possibly reflecting a strategy focused on performance optimization. Cluster 3 exhibits the most negative slack, suggesting less stringent timing constraints or less optimized designs. Finally, Fig. \ref{partyDistributions}(c) shows the distribution of Area, revealing that Cluster 2 generally occupies larger areas, which might indicate a design strategy prioritizing complexity over compactness, while Cluster 3 occupies intermediate areas, balancing complexity and compactness. In summary, Cluster 1 focuses on energy-efficient designs (e.g., IoT devices), Cluster 2 prioritizes high-complexity designs (e.g., data center accelerators), and Cluster 3 prioritizes timing-critical applications (e.g., high-throughput systems).

\par \textbf{Analysis of Non-IID-ness Across Clusters.} To investigate the degree of Non-Independent and Identically Distributedness (non-IID-ness) across the three sub-datasets, we analyze the inter-cluster differences in Area, Slack, and Total Power distributions. Such differences justify FL by highlighting the benefits of sharing model updates \cite{mcmahan2017communication}.

\par
\begin{figure}
 \centering
 \includegraphics[width=0.88\linewidth]{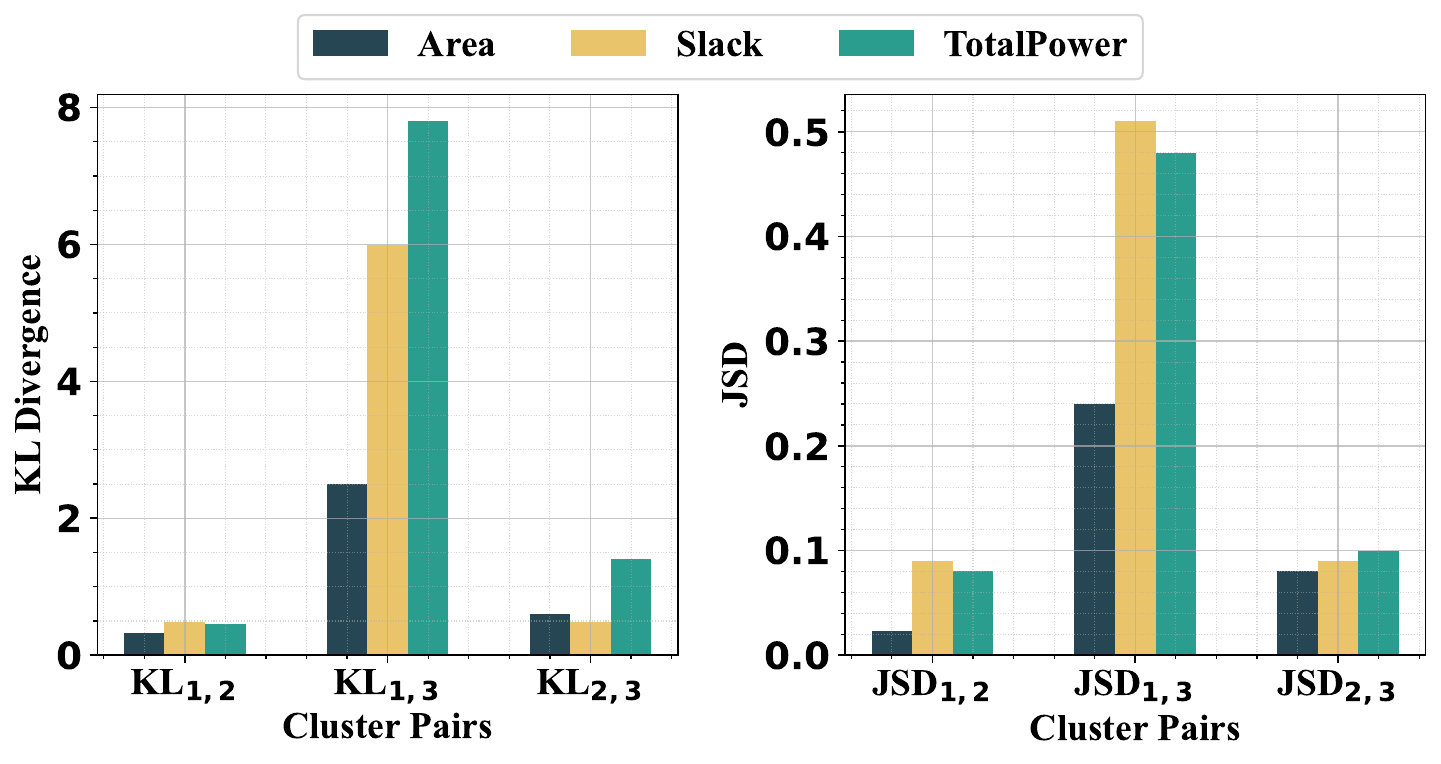}\vspace{-1.25em}
 \caption{Divergence measures across clusters (a) KL-Divergence between clusters, (b) Jensen-Shannon Divergence between clusters.}
 \label{nonIIDNess}\vspace{-1.25em}
\end{figure}

 We characterize the inter-cluster differences in the distributions using two measures: KL-Divergence \cite{kullback1951information} and Jensen-Shannon Divergence (JSD), as shown in Fig.\ref{nonIIDNess}(a) and Fig.\ref{nonIIDNess}(b), respectively. The subscripts in the cluster pairs denote the pairwise comparisons between the corresponding clusters. Both measures indicate that Total Power exhibits the highest divergence, particularly between Clusters 1 and 3, reflecting significant non-IID-ness in this metric. Slack shows moderate divergence, with notable differences between Clusters 1 and 3 as well as Clusters 2 and 3. Area generally exhibits the least divergence, indicating higher similarity across the clusters for this metric. KL-Divergence, being more sensitive, highlights larger differences, particularly for Total Power, while JSD provides a more balanced and bounded interpretation of the divergence between distributions. These results highlight significant non-IID characteristics across the three client datasets, highlighting the benefits of coefficient sharing in the FL paradigm.

\subsection{The \textit{Chip@k} Metric for Chip Design Evaluation}
\par Evaluating the quality of generated chip designs requires a metric that accounts for multiple, often conflicting design objectives. To address this need, we introduce the \textit{Chip@k} metric, which quantifies model performance stochastically by measuring the proportion of generated designs that meet statistical acceptability criteria across these metrics.

\par Each design is evaluated using the \textit{Three Sigma Rule}, which defines reasonable bounds for deviations from ground truth values based on standard deviations within the dataset. Specifically, for a dataset \( D = \{(x_i, y_i), i \in [1, t]\} \), where \(t\) is the number of designs, each design \(y_i\) is associated with three metrics: \({y_i}^{\text{AREA}}\), \({y_i}^{\text{SLACK}}\), and \({y_i}^{\text{POWER}}\). Given a design description \(x_i\), the model generates a design \(y'_i\), and the deviation for a metric, such as Area, is computed as \({\delta_i}^{\text{AREA}} = {y'_i}^{\text{AREA}} - {y_i}^{\text{AREA}}\). It is noted that the lower $\delta$ is the better, and a negative $\delta$ means a design with even less area compared to the desired ground truth. To determine acceptance, the distribution \(\mathcal{D}^{\text{AREA}}\) of ground truth values \({y_i}^{\text{AREA}}\) and its standard deviation \(\sigma^{\text{AREA}}\) are computed. The generated value \({y'_i}^{\text{AREA}}\) is accepted if \({\delta_i}^{\text{AREA}} < \sigma^{\text{AREA}}\). 

\par For a normal distribution, approximately \(68\%\), \(95\%\), and \(99.7\%\) of the data lie within one, two, and three standard deviations, respectively, of the mean \cite{pukelsheim1994three}. Therefore, we guarantee that $Pr({\delta_i}^{AREA} < \sigma^{AREA}) \approx 2.1\% + 13.6 \% + 68.2 \% = 83.9\%$. We apply the same process to the other two metrics.


\par The \textit{Chip@k} metric measures the expected probability that at least one of the top-\(k\) generated designs satisfies all acceptance criteria. It is defined as:
\begin{equation}
\label{eq:chip_at_k}
\text{Chip@}k = \mathbb{E}_{d \sim \mathcal{D}} \left[ 1 - \frac{\binom{n - c_d}{k}}{\binom{n}{k}} \right],
\end{equation}
\noindent where \(n\) is the total number of generated candidates for a design description \(d\), \(c_d\) is the number of candidates passing the Three Sigma Rule for all three PPA metrics, and \(\mathcal{D}\) denotes the distribution over design descriptions.

\subsection{Client-Side Fine-Tuning}
\par \texttt{FedChip} uses the cross-entropy loss to fine-tune the model for Verilog code generation. This loss measures the discrepancy between the ground-truth Verilog code and the code by generated comparing their token probability distributions.
\begin{equation}
L = -\frac{1}{N} \sum_{i=1}^N \sum_{j=1}^C y_{ij} \log(p_{ij}),
\label{eq2}
\end{equation}
\noindent where \( N \) is the batch size, \( C \) is the vocabulary size, \( y_{ij} \) is the ground-truth token (one-hot encoded), and \( p_{ij} \) is the predicted token probability. This loss penalizes incorrect tokens to promote accurate Verilog generation.

\begin{figure}[h]
 \centering \includegraphics[width=0.555\linewidth]{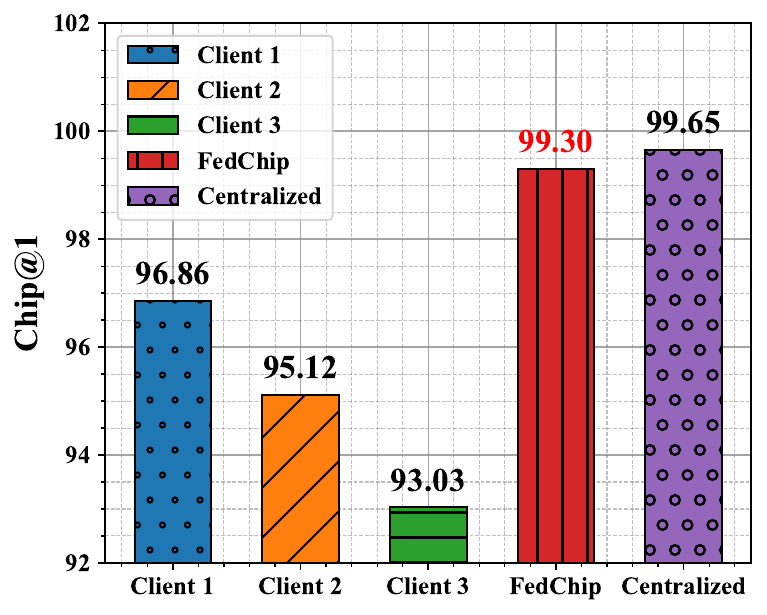}
 \vspace{-1.2em}
 \caption{Comparison of performance metrics between centralized and federated fine-tuning scenarios.} 
 \label{perfoamnceEval}\vspace{-1.35em}
\end{figure}

\par To enable resource-constrained clients to participate in fine-tuning, \texttt{FedChip} integrates LoRA. LoRA efficiently updates a small subset of model parameters by injecting low-rank decomposed matrices into the model's weight updates, significantly reducing memory and computational overhead during training. This approach allows each client to fine-tune the model locally without requiring access to the full model's parameters, preserving privacy and enabling scalability. While effective, this setup does not yet address communication overhead or adapt LoRA ranks to client capabilities. Both are promising directions we leave for future work.

\section{Evaluation}
\label{section5}

\par \textbf{The Setup.} Our experiments are based on a total of 30k sample designs, as described in Section \ref{section3}. We assume an FL system for hardware generation as shown in Fig. \ref{fig0} with 3 parties (clients). Each party represents an independent hardware design manufacturer with distinct design priorities. We use Mistral 7B as the underlying LLM, and the experiments are conducted on appropriate hardware infrastructure. At each client, we pick 500 data points as a test set and use the remaining datasets of the same sub-dataset for training. The evaluation focuses on three key PPA quality metrics.

\begin{figure*}[t]
 \centering
 \resizebox{0.83\textwidth}{!}{%
 \begin{tabular}{ccc}
 \includegraphics[width=0.33\textwidth]{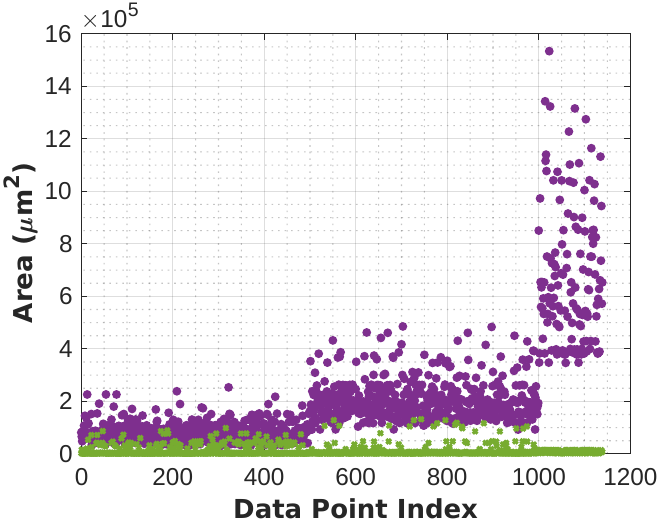} &
 \includegraphics[width=0.33\textwidth]{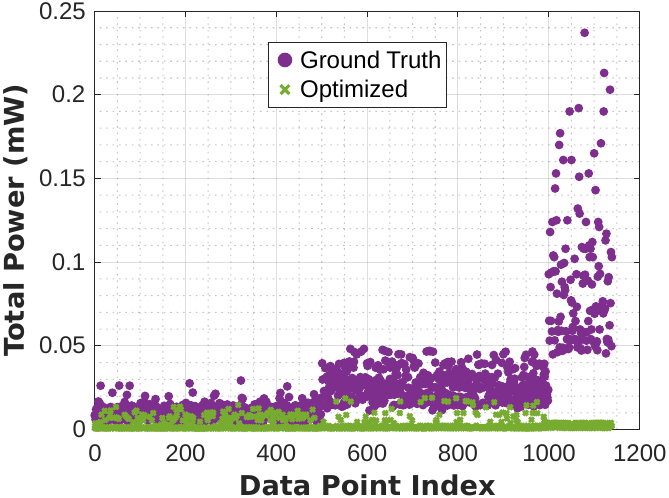} &
 \includegraphics[width=0.33\textwidth]{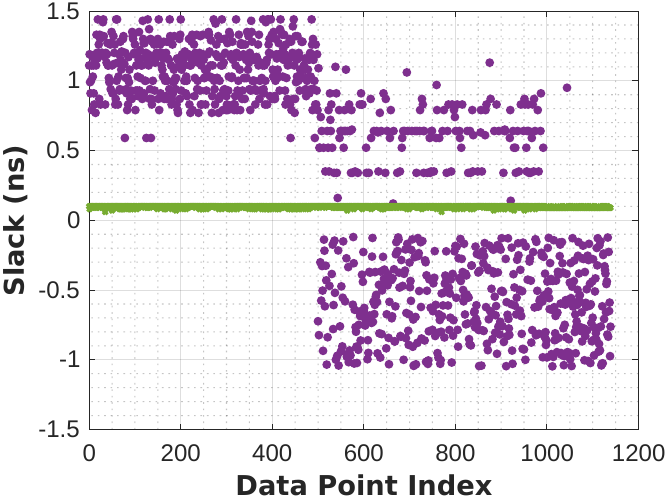} \\
  (a) & (b) & (c)
 \end{tabular}
 }\vspace{-0.55em}
 \caption{Comparison of FL fine-tuned and aggregated model metrics vs. ground truth for the area, total power, and slack in (a) through (c), respectively.}
 \label{FLvsGT}\vspace{-1.25em}
\end{figure*}

\par \textbf{Fine-Tuning Specifications.} We conduct experiments on \texttt{FedChip} using the FedAvg \cite{mcmahan2017communication} setting within the OpenFedLLM \cite{ye2024openfedllmtraininglargelanguage} framework. To improve computational efficiency, we quantize the model to int8 during training while preserving critical operations in floating point. The fine-tuning process uses a maximum sequence length of 1024, a batch size of 16, and a learning rate of 2e-5. For LoRA \cite{hu2021loralowrankadaptationlarge}, we configure a rank of 8 and a scaling factor ($\alpha$) of 16. Each client trains locally on their dataset using the AdamW optimizer \cite{loshchilov2019decoupledweightdecayregularization}, with instruction formatting guided by the CodeAlpaca \cite{alpaca} template. For centralized training, we combine all client datasets into one and use the same parameters to ensure a fair comparison with the federated setting. This allows a systematic evaluation of the benefits of \texttt{FedChip}'s federated approach over centralized training.

\subsection{Performance Evaluation}

\par The evaluation compares three scenarios: (i) a centralized setting where all data from the three parties is aggregated and used to train a single model, (ii) the \texttt{FedChip} framework where federated fine-tuning is applied using the \textit{Chip@k} metric, enabling collaborative training without sharing raw data, and (iii) independent training where each party trains its model locally using only its own data. This setup allows for a comprehensive assessment of the trade-offs between privacy, collaboration, and model performance.

\par Fig.~\ref{perfoamnceEval} illustrates the effectiveness of \texttt{FedChip} in achieving a balance between performance and privacy in hardware design. \texttt{FedChip} achieves near-centralized performance under the \textit{Chip@1} metric, with only a marginal drop of \(0.35\%\) compared to the centralized setting (\(99.65\%\) vs. \(99.30\%\)). Additionally, \texttt{FedChip} substantially outperforms the independently trained models of Client 1, Client 2, and Client 3, whose performances are \(96.86\%\), \(95.12\%\), and \(93.03\%\), respectively. These results highlight \texttt{FedChip}'s capability to significantly enhance model performance across diverse datasets while ensuring data privacy and avoiding the need for centralized data collection.


\par Fig.~\ref{FLvsGT} compares the PPA metrics generated by the FL fine-tuned model to the ground truth values. Blue dots represent the ground truth metrics, and red crosses indicate the model's optimized outputs. (a) shows the comparison for Area, where the optimized values demonstrate significantly less variability and cluster near the lower bounds, highlighting the model's efficiency in minimizing area. (b) compares Total Power, with the optimized results showing reduced variability and lower power levels compared to the ground truth. (c) evaluates Slack, where the ground truth values show high variability, including negative values, while the optimized results remain consistently positive and tightly grouped, ensuring timing constraints are met. These results highlight the model's ability to produce designs that meet constraints with lower variability and comparable or improved performance.

\subsection{Ablation Study}

\par To evaluate the added benefit of LLM fine-tuning with \textit{FedChip}, we conduct an experiment using four high-end LLMs: OpenAI's GPT-4o \cite{openai2024gpt4o}, OpenAI's GPT-o1 \cite{openai2024gpto1}, Google's Gemini Advanced \cite{gemini2024advanced}, and Anthropic's Claude 3.5 Sonnet \cite{anthropic2024claude3.5}. These LLMs are used in their vanilla state (i.e., without fine-tuning). For a fair comparison, we augment the verbal descriptions in original prompts with a 1-shot example from the dataset to guide the LLMs toward accurate generation.

\par
\begin{figure}[h]
\centering \includegraphics[width=0.555\linewidth]{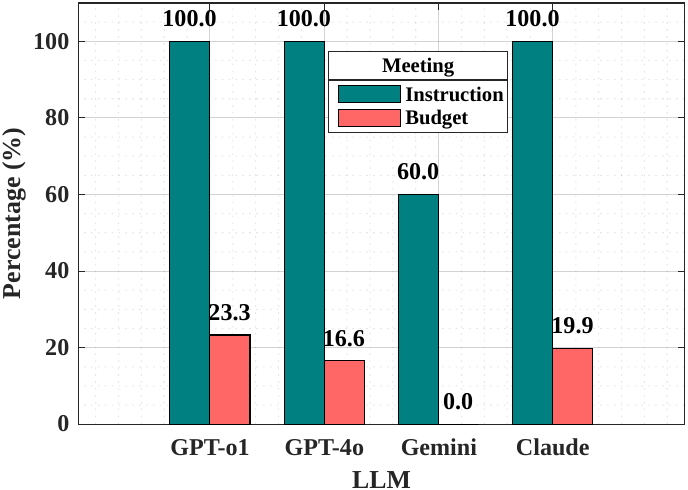}\vspace{-1em}
\caption{Ablation study of vanilla high-end LLMs with 1-shot learning, evaluating code generation accuracy and PPA constraint compliance.}
\label{HighLLM}\vspace{-1em}
\end{figure}

 The performance is evaluated based on two metrics: \encircle{1} the ability to generate synthesizable design code that matches the verbal description and \encircle{2} the compliance of the generated designs with desired PPA constraints. The results, shown in Fig.~\ref{HighLLM}, highlight that while these LLMs can generate synthesizable code, they fail to adhere to the design constraints. This underscores the necessity of fine-tuning with \textit{FedChip} to achieve both design accuracy and constraint compliance.
\par These findings highlight the importance of domain-specific fine-tuning for hardware optimization. \textit{FedChip} enhances constraint adherence and supports scalable, collaborative refinement across diverse non-IID datasets. This has significant implications for design automation, emphasizing the importance of tailored LLM training to meet industry-specific needs. As seen, \textit{FedChip} outperforms the highest performing vanilla LLM (GPT-O1) By around 77\% 

\section{Conclusion}
\par This paper presents \texttt{APTPU-Gen}, a comprehensive dataset of 30k verified hardware design configurations annotated with detailed PPA metrics, specifically generated to evaluate LLM-based hardware optimization frameworks. The dataset is clustered to reflect the unique focuses of multiple hardware manufacturers, exhibiting high non-IIDness to capture diverse design strategies and priorities, enabling realistic multi-party FL. Using \texttt{APTPU-Gen}, we propose \texttt{FedChip}, a federated fine-tuning framework for automated hardware design generation with LLMs. \texttt{FedChip} enables collaborative model enhancement while preserving data privacy and addressing conflicting design objectives. To assess goal fulfillment in generated designs, we introduce the \textit{Chip@k} metric, which evaluates critical PPA parameters such as Total PPA. Experimental results show that \texttt{FedChip} significantly improves design quality over centralized training and vanilla LLMs. Future work includes developing privacy guarantees, advanced loss functions for variable design generations, and more efficient update mechanisms.

\section*{Acknowledgment}
\small \vspace{-0.5em}
This work is supported in part by the National Science Foundation (NSF) under grant no. 2216772, 2228028, 2409697 and Semiconductor Research Corporation (SRC).\vspace{-1.05em}

\bibliographystyle{IEEEtran}
\bibliography{references}

\begin{thebibliography}{10}
\providecommand{\url}[1]{#1}
\csname url@samestyle\endcsname
\providecommand{\newblock}{\relax}
\providecommand{\bibinfo}[2]{#2}
\providecommand{\BIBentrySTDinterwordspacing}{\spaceskip=0pt\relax}
\providecommand{\BIBentryALTinterwordstretchfactor}{4}
\providecommand{\BIBentryALTinterwordspacing}{\spaceskip=\fontdimen2\font plus
\BIBentryALTinterwordstretchfactor\fontdimen3\font minus \fontdimen4\font\relax}
\providecommand{\BIBforeignlanguage}[2]{{%
\expandafter\ifx\csname l@#1\endcsname\relax
\typeout{** WARNING: IEEEtran.bst: No hyphenation pattern has been}%
\typeout{** loaded for the language `#1'. Using the pattern for}%
\typeout{** the default language instead.}%
\else
\language=\csname l@#1\endcsname
\fi
#2}}
\providecommand{\BIBdecl}{\relax}
\BIBdecl

\bibitem{skyquest2023aihardware}
\BIBentryALTinterwordspacing
{SkyQuest Technology}, ``Artificial intelligence (ai) hardware market to exceed usd 84.9 billion by 2031: Skyquest technology,'' 2023, accessed: 2024-10-31. [Online]. Available: \url{https://www.prnewswire.com}
\BIBentrySTDinterwordspacing

\bibitem{ren2023survey}
W.-Q. Ren \emph{et~al.}, ``A survey on collaborative dnn inference for edge intelligence,'' \emph{Machine Intelligence Research}, vol.~20, no.~3, pp. 370--395, 2023.

\bibitem{genc2021gemmini}
H.~Genc \emph{et~al.}, ``Gemmini: Enabling systematic deep-learning architecture evaluation via full-stack integration,'' in \emph{DAC}.\hskip 1em plus 0.5em minus 0.4em\relax IEEE, 2021, pp. 769--774.

\bibitem{thakur2023verigen}
S.~Thakur, B.~Ahmad, H.~Pearce, B.~Tan, B.~Dolan-Gavitt, R.~Karri, and S.~Garg, ``Verigen: A large language model for verilog code generation,'' \emph{ACM Transactions on Design Automation of Electronic Systems}, vol.~29, no.~3, pp. 1--31, 2024.

\bibitem{he2023chateda}
H.~Wu, Z.~He, X.~Zhang, X.~Yao, S.~Zheng, H.~Zheng, and B.~Yu, ``Chateda: A large language model powered autonomous agent for eda,'' \emph{IEEE Transactions on Computer-Aided Design of Integrated Circuits and Systems}, 2024.

\bibitem{chang2023chipgpt}
K.~Chang, Y.~Wang, H.~Ren, M.~Wang, S.~Liang, Y.~Han, H.~Li, and X.~Li, ``Chipgpt: How far are we from natural language hardware design,'' \emph{arXiv preprint arXiv:2305.14019}, 2023.

\bibitem{thakur2023autochip}
S.~Thakur \emph{et~al.}, ``Autochip: Automating hdl generation using llm feedback,'' \emph{arXiv preprint arXiv:2311.04887}, 2023.

\bibitem{10323953}
Y.~Fu, Y.~Zhang, Z.~Yu, S.~Li, Z.~Ye, C.~Li, C.~Wan, and Y.~C. Lin, ``Gpt4aigchip: Towards next-generation ai accelerator design automation via large language models,'' in \emph{2023 IEEE/ACM International Conference on Computer Aided Design (ICCAD)}.\hskip 1em plus 0.5em minus 0.4em\relax IEEE, 2023, pp. 1--9.

\bibitem{vungarala2024sa}
D.~Vungarala, M.~Nazzal, M.~Morsali, C.~Zhang, A.~Ghosh, A.~Khreishah, and S.~Angizi, ``Sa-ds: A dataset for large language model-driven ai accelerator design generation,'' \emph{arXiv e-prints}, pp. arXiv--2404, 2024.

\bibitem{openai2024gpt4o}
{OpenAI}, ``Gpt-4o: Openai's advanced generative language model,'' Available online: \url{https://openai.com}, 2024, accessed: 2024-11-18.

\bibitem{anthropic2024claude3.5}
{Anthropic}, ``Claude 3.5 sonnet: Anthropic's advanced language model,'' Available online: \url{https://www.anthropic.com}, 2024, accessed: 2024-11-18.

\bibitem{bommasani2021opportunities}
R.~Bommasani, D.~A. Hudson, E.~Adeli, R.~Altman, S.~Arora, S.~von Arx, M.~S. Bernstein, J.~Bohg, A.~Bosselut, E.~Brunskill \emph{et~al.}, ``On the opportunities and risks of foundation models,'' \emph{arXiv preprint arXiv:2108.07258}, 2021.

\bibitem{xu2024llm}
K.~Xu, R.~Qiu, Z.~Zhao, G.~L. Zhang, U.~Schlichtmann, and B.~Li, ``Llm-aided efficient hardware design automation,'' \emph{arXiv preprint arXiv:2410.18582}, 2024.

\bibitem{chen2024dawn}
L.~Chen, Y.~Chen, Z.~Chu, W.~Fang, T.-Y. Ho, R.~Huang, Y.~Huang, S.~Khan, M.~Li, X.~Li \emph{et~al.}, ``The dawn of ai-native eda: Opportunities and challenges of large circuit models,'' \emph{arXiv preprint arXiv:2403.07257}, 2024.

\bibitem{arstechnica2022nvidiacyberattack}
\BIBentryALTinterwordspacing
D.~Goodin, ``Cybercriminals who breached nvidia issue one of the most unusual demands ever,'' \emph{Ars Technica}, March 2022, accessed: 2024-10-31. [Online]. Available: \url{https://arstechnica.com}
\BIBentrySTDinterwordspacing

\bibitem{davis2021fast}
W.~R. Davis, P.~Franzon, L.~Francisco, B.~Huggins, and R.~Jain, ``Fast and accurate ppa modeling with transfer learning,'' in \emph{2021 IEEE/ACM International Conference On Computer Aided Design (ICCAD)}.\hskip 1em plus 0.5em minus 0.4em\relax IEEE, 2021, pp. 1--8.

\bibitem{yang2019federated}
Q.~Yang, Y.~Liu, T.~Chen, and Y.~Tong, ``Federated machine learning: Concept and applications,'' \emph{ACM Transactions on Intelligent Systems and Technology (TIST)}, vol.~10, no.~2, pp. 1--19, 2019.

\bibitem{bonawitz2019towards}
K.~Bonawitz, ``Towards federated learning at scale: Syste m design,'' \emph{arXiv preprint arXiv:1902.01046}, 2019.

\bibitem{mcmahan2017communication}
B.~McMahan, E.~Moore, D.~Ramage, S.~Hampson, and B.~A. y~Arcas, ``Communication-efficient learning of deep networks from decentralized data,'' in \emph{Artificial intelligence and statistics}.\hskip 1em plus 0.5em minus 0.4em\relax PMLR, 2017, pp. 1273--1282.

\bibitem{bai2024federated}
J.~Bai, D.~Chen, B.~Qian, L.~Yao, and Y.~Li, ``Federated fine-tuning of large language models under heterogeneous language tasks and client resources,'' \emph{arXiv preprint arXiv:2402.11505}, 2024.

\bibitem{kuang2024federatedscope}
W.~Kuang, B.~Qian, Z.~Li, D.~Chen, D.~Gao, X.~Pan, Y.~Xie, Y.~Li, B.~Ding, and J.~Zhou, ``Federatedscope-llm: A comprehensive package for fine-tuning large language models in federated learning,'' in \emph{Proceedings of the 30th ACM SIGKDD Conference on Knowledge Discovery and Data Mining}, 2024, pp. 5260--5271.

\bibitem{kuo2024federated}
K.~Kuo, A.~Raje, K.~Rajesh, and V.~Smith, ``Federated lora with sparse communication,'' \emph{arXiv preprint arXiv:2406.05233}, 2024.

\bibitem{shi2020learned}
Z.~Shi, C.~Sakhuja, M.~Hashemi, K.~Swersky, and C.~Lin, ``Learned hardware/software co-design of neural accelerators,'' \emph{arXiv preprint arXiv:2010.02075}, 2020.

\bibitem{dai2022can}
D.~Dai, Y.~Sun, L.~Dong, Y.~Hao, S.~Ma, Z.~Sui, and F.~Wei, ``Why can gpt learn in-context? language models implicitly perform gradient descent as meta-optimizers,'' \emph{arXiv preprint arXiv:2212.10559}, 2022.

\bibitem{brown2020language}
T.~B. Brown, ``Language models are few-shot learners,'' \emph{arXiv preprint arXiv:2005.14165}, 2020.

\bibitem{devlin2019bertpretrainingdeepbidirectional}
\BIBentryALTinterwordspacing
J.~Devlin, M.-W. Chang, K.~Lee, and K.~Toutanova, ``Bert: Pre-training of deep bidirectional transformers for language understanding,'' 2019. [Online]. Available: \url{https://arxiv.org/abs/1810.04805}
\BIBentrySTDinterwordspacing

\bibitem{liu2019roberta}
Y.~Liu, ``Roberta: A robustly optimized bert pretraining approach,'' \emph{arXiv preprint arXiv:1907.11692}, vol. 364, 2019.

\bibitem{hu2021lora}
E.~J. Hu, Y.~Shen, P.~Wallis, Z.~Allen-Zhu, Y.~Li, S.~Wang, L.~Wang, and W.~Chen, ``Lora: Low-rank adaptation of large language models,'' \emph{arXiv preprint arXiv:2106.09685}, 2021.

\bibitem{dettmers2024qlora}
T.~Dettmers, A.~Pagnoni, A.~Holtzman, and L.~Zettlemoyer, ``Qlora: Efficient finetuning of quantized llms,'' \emph{Advances in Neural Information Processing Systems}, vol.~36, 2024.

\bibitem{zhang2022opt}
S.~Zhang, S.~Roller, N.~Goyal, M.~Artetxe, M.~Chen, S.~Chen, C.~Dewan, M.~Diab, X.~Li, X.~V. Lin \emph{et~al.}, ``Opt: Open pre-trained transformer language models,'' \emph{arXiv preprint arXiv:2205.01068}, 2022.

\bibitem{TPU-micro2018}
N.~Jouppi, C.~Young, N.~Patil, and D.~Patterson, ``Motivation for and evaluation of the first tensor processing unit,'' \emph{IEEE Micro}, vol.~38, no.~3, pp. 10--19, 2018.

\bibitem{9901385}
M.~E. Elbtity, P.~S. Chandarana, B.~Reidy, J.~K. Eshraghian, and R.~Zand, ``Aptpu: Approximate computing based tensor processing unit,'' \emph{IEEE Transactions on Circuits and Systems I: Regular Papers}, vol.~69, no.~12, pp. 5135--5146, 2022.

\bibitem{OPENROAD}
\BIBentryALTinterwordspacing
(2018) Openroad. [Online]. Available: \url{https://github.com/The-OpenROAD-Project/OpenROAD}
\BIBentrySTDinterwordspacing

\bibitem{lloyd1982least}
S.~Lloyd, ``Least squares quantization in pcm,'' \emph{IEEE transactions on information theory}, vol.~28, no.~2, pp. 129--137, 1982.

\bibitem{minka2000estimating}
T.~Minka, ``Estimating a dirichlet distribution,'' 2000.

\bibitem{kullback1951information}
S.~Kullback and R.~A. Leibler, ``On information and sufficiency,'' \emph{The annals of mathematical statistics}, vol.~22, no.~1, pp. 79--86, 1951.

\bibitem{pukelsheim1994three}
F.~Pukelsheim, ``The three sigma rule,'' \emph{The American Statistician}, vol.~48, no.~2, pp. 88--91, 1994.

\bibitem{ye2024openfedllmtraininglargelanguage}
\BIBentryALTinterwordspacing
R.~Ye, W.~Wang, J.~Chai, D.~Li, Z.~Li, Y.~Xu, Y.~Du, Y.~Wang, and S.~Chen, ``Openfedllm: Training large language models on decentralized private data via federated learning,'' 2024. [Online]. Available: \url{https://arxiv.org/abs/2402.06954}
\BIBentrySTDinterwordspacing

\bibitem{hu2021loralowrankadaptationlarge}
\BIBentryALTinterwordspacing
E.~J. Hu, Y.~Shen, P.~Wallis, Z.~Allen-Zhu, Y.~Li, S.~Wang, L.~Wang, and W.~Chen, ``Lora: Low-rank adaptation of large language models,'' 2021. [Online]. Available: \url{https://arxiv.org/abs/2106.09685}
\BIBentrySTDinterwordspacing

\bibitem{loshchilov2019decoupledweightdecayregularization}
\BIBentryALTinterwordspacing
I.~Loshchilov and F.~Hutter, ``Decoupled weight decay regularization,'' 2019. [Online]. Available: \url{https://arxiv.org/abs/1711.05101}
\BIBentrySTDinterwordspacing

\bibitem{alpaca}
R.~Taori, I.~Gulrajani, T.~Zhang, Y.~Dubois, X.~Li, C.~Guestrin, P.~Liang, and T.~B. Hashimoto, ``Stanford alpaca: An instruction-following llama model,'' \url{https://github.com/tatsulab/stanford_alpaca}, 2023.

\bibitem{openai2024gpto1}
{OpenAI}, ``Gpt-o1: Specialized generative pre-trained transformer for domain applications,'' Available online: \url{https://openai.com}, 2024, accessed: 2024-11-18.

\bibitem{gemini2024advanced}
{Gemini AI Lab}, ``Gemini advanced: High-performance large language model,'' Available online: \url{https://www.gemini-ai.com}, 2024, accessed: 2024-11-18.

\end{thebibliography}

\end{document}